\documentclass[12pt]{iopart}
\usepackage{graphicx}
\begin{document}
\title[Tracking tumor evolution via the prostate marker PSA]{Tracking tumor evolution via the prostate marker PSA: An individual post-operative study}

\author{Mehmet Erbudak$^{1,2}$ and Ay\c se Erzan$^{3}$\footnote{Permanent address:
Department of Physics, Faculty of Letters and Sciences, Istanbul Technical University, Maslak, 34\,469 Istanbul, Turkey}}

\address{$^1$Laboratory for Solid State Physics, ETH Zurich, CH-8093 Zurich, Switzerland}
\address{$^2$Physics Department, Bo\u gazi\c ci University, 34\,342 Bebek, Istanbul, Turkey}
%\address{$^3$Physics Department,  Istanbul Technical University, Istanbul, Turkey}
\address{$^3$Department of Physics, Faculty of Letters and Sciences, Akdeniz University, Dumlupinar Bulvari, 07\,058 Antalya, Turkey}

\eads{\mailto {erbudak@phys.ethz.ch}, \mailto {erzan@itu.edu.tr}}

\begin{abstract}
The progress of the prostate-specific antigen after radical prostatectomy is observed for a patient in order to extract information on the growth mode of the tumor cells. An initial fast-growth mode goes over to a slower power-law regime within two years of surgery. We argue that such studies may help determine the appropriate time window for subsequent therapies in order to increase the life expectancy of the patient.
\end{abstract}

\pacs{87.18.Hf, 87.55.-x, 89.75.Da }
\vspace{2pc}
\noindent{\it Keywords\/}: tumor growth, Eden model, scaling behavior, Gompertzian growth, percolation, prostate cancer, prostate marker PSA, post-operative treatment strategies

\submitto{\PMB}
\maketitle

\section{Introduction}
The cancer of the prostate gland is one of the most frequently diagnosed male illnesses that may lead to the death of individuals older than 50. There are two major causes for the dramatic increase in the number of men diagnosed with prostate cancer every year: i) a straight-forward blood test became available measuring a glycoprotein called prostate-specific antigen (PSA) to detect prostate carcinoma, ii) by virtue of advancing medical techniques, the increasing life expectancy exposes a higher male population to the disease.

The PSA test is a statistical decision-making tool. While it detects true-positive cases with ease, the false-negative outcome is not nil and may have a lethal consequence. Hence, a biopsy of the gland provides a ``second opinion". In the case of positive outcome of the biopsy, statistical predictions are used to assess the status of the tumor, its classification, and the extent of its progression. If the result of the PSA test turns out to be a false alarm, the inconvenience of the biopsy is the consequence.

At an early stage of the cancer growth with a localized tumor, a radical removal of the prostate gland is preferred for younger patients. It is generally believed that a conformal radiation therapy may be as effective, however, in the case of elder individuals. For cases of progressed tumor, 
palliative measures are preferred including hormonal therapy to retard the growth of the prostate cancer. Findings of the last few years based on long-term statistics suggest a longer life expectancy for patients with postoperative radiotherapy that follows (within 6 months of) the radical prostate surgery (see, e.g., Cozzarini \etal 2004, Bolla \etal 2005). Albeit known to cause adverse side effects, this strategy may certainly prove profitable for patients with some higher risks, like prostate capsule perforation or involvement of seminal vesicles or lymph-nodes.

Different alternatives for treatment after radical removal of the prostate gland are currently under debate if the PSA values rise after the surgery. Commonly, a standard radiotherapy is applied after the PSA level reaches a predetermined threshold value. However, a wait-and-watch method may cause loss of valuable time and miss the relevant moment of action regardless of how low the threshold value is set.

It has already been suggested by many authors over the years (Schabel 1969, Br\'u \etal 2003, Kohandel \etal 2007, these references are only a sample and do not in any way claim to be exhaustive), that the entire course of growth of the tumors offers important information regarding the clinical strategies to be followed. Here we would like to point out a number of features of the the post-operative temporal progression of PSA values which could help make a clearer decision about the strategic time window for combined therapies on these cells.

The purpose of this work is three-fold:
a) point to the possibility of detecting  fast (exponential) growth of the PSA scores, much before an arbitrary threshold value is reached, thus gaining time in making decisions regarding therapies to be followed, 
b) re-analyze PSA data in a way which reveals a sharp crossover from exponential to power-law growth c) propose a simple model to explain the crossover to slower (power-law) growth. 

The second, slower growth regime, we argue, arises due to the coalescence or ``condensation" (Torkington 1983) of freely dividing cancerous cells to form one or more compact tumors, with growth essentially confined to the edges or the surface (Br\'u \etal 2003, Kansal \etal 2000, Delsanto \etal 2004, Kohandal \etal 2007). It has been pointed out that at this stage ``sensitivity to anti-metabolic drugs decreases,$\ldots$ (since the fraction of) tumor cells that are in the cell division cycle decreases" (Schabel 1969).  It is only subsequent to this crossover that processes such as angiogenesis (Kohandel \etal 2007) can come into play. 

\section {Results}

\begin{table}[!b]
\caption{The PSA values and the dates of measurement after the operation in March 2003 as well as the time elapsed thereafter in months. The errors in determining the PSA values are $\pm 0.002~\mu$g/l, while the time elapsed  after the operation  may err by at most  $\pm 1$ week.}

\begin{indented}
\item[]
\begin{tabular}{@{}|c|c|c|}
\br
\; Date of the \;& Time after the & \; PSA score \;\\ PSA test & operation (months) & ($\mu$g/l) \\
\mr
Aug. 2003 & 5 & 0.006\\
Feb. 2004 & 11 & 0.012\\
Aug. 2004 & 17 & 0.019\\
Feb. 2005 & 23 & 0.037\\
Aug. 2005 & 29 & 0.044\\
Feb. 2006 & 35 & 0.099\\
Aug. 2006 & 41 & 0.170\\
Nov. 2006 & 44 & 0.144\\
Feb. 2007 & 47 & 0.168\\
May 2007 & 50 & 0.294\\
Jul. 2007 & 52 & 0.262\\
Feb. 2008 & 59 & 0.295\\
Aug. 2008 & 65 & 0.485\\
\br
\end{tabular}
\end{indented}
\end{table}

\begin{figure}

\includegraphics[width=0.67\columnwidth]{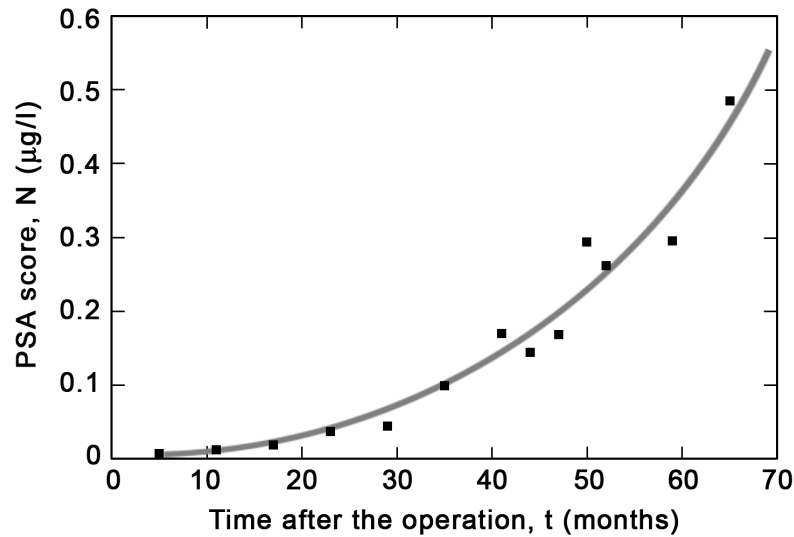}
\caption{PSA values in $\mu$g/l are displayed as a function of measurement time in months after the surgery.}
\label{linear}
\end{figure}

After a radical prostatovesiculectomy (pT2c N0 M0 G2, Gleason $3+4=7$) applied to one of us (ME), PSA values have been determined using constant laboratory conditions at time intervals of initially 6 months. These values are presented in Table 1. The estimated error for the dates of measurement is $\pm 1$ week, while the PSA values are determined with the state-of-the-art precision, with error bars of the order of $\pm 0.002~\mu$g/l. In Fig.~\ref{linear} we plot all the values listed in the table as a function of time in months after the surgery. The temporal increase of the PSA scores has a characteristic ``U" shape, i.e., a shallow increase during the first two years and a steep rise in the last two. In current practice, each datum point at a particular time is used by the physician to assess the patient's health condition and to decide upon further action. Here we show that the functional form of this curve can be determined at an early stage which provides crucially important information on the stage of progression of the disease.

\begin{figure}[t!]
\includegraphics[width=1.0\columnwidth]{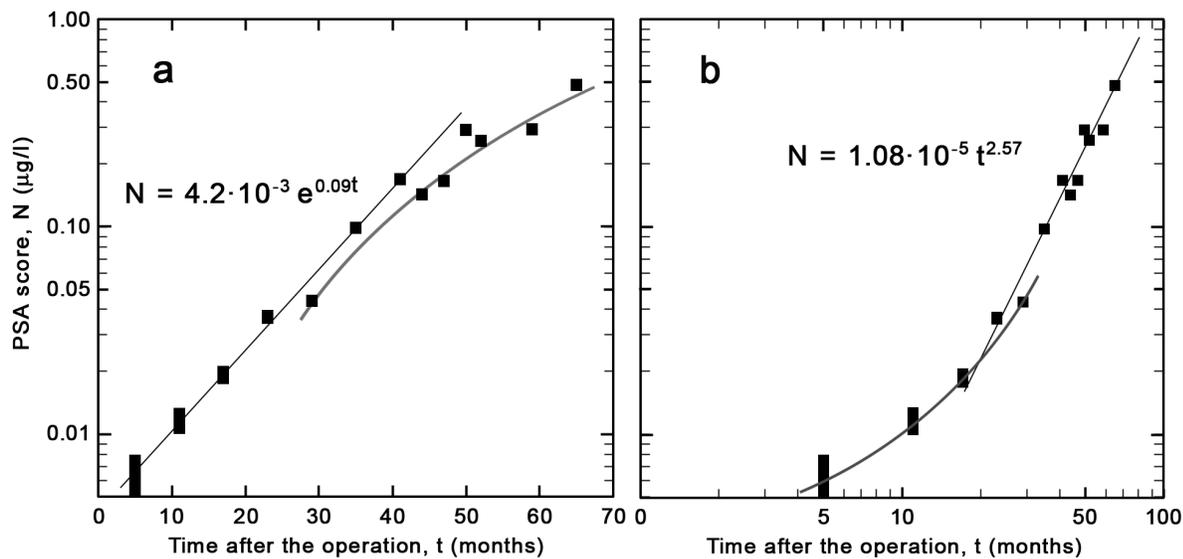}
\caption{PSA values plotted on a logarithmic scale against a) linear time and b) against time on a logarithmic scale.  Note that the straight line fit in panel (a) signals exponential growth, while in panel (b), ``power law" growth (see text). The vertical size of the first four points indicates the estimated error, while in the subsequent points the error bars are smaller than the size of the plotted points.} 
\label{log}
\end{figure}

In order to detect the kinetics of cell growth underlying the PSA progression, we present in Fig.~\ref{log}(a) the same PSA values as a function of time, but plotted on a logarithmic scale. We observe a linear increase starting at the time of surgery until about 40 months thereafter. Analytically this corresponds to ``exponential growth," with the functional form $\sim \exp (at)$, where $a$ is a constant rate of reproduction. Later values remain below the straight line, implying a different, slower growth law.

We examine more closely the crossover from exponential behavior in the real PSA data given in Table 1  by re-plotting the values against the number of months after the surgery, this time with both axes in the logarithmic scale. In this log-log plot,  a power-law behavior in time (i.e., having the functional form $\sim t^u$) is represented by a straight line. Fig.~\ref{log}(b) shows the results and indicates that the crossover observed in Fig.~\ref{log}(a) at about two years after the surgery is remarkably sharp, rather than being a gradual slowing down. From this point on, up to the last measured value, the PSA values grow as a power of the time elapsed  after the crossover point.

\section {Procedures}

\subsection{Growth modes}

We assume that the PSA value in $\mu$g/l is linearly proportional to the number $N$ of carcinoma cells. We also assume that each cell divides with a constant rate $p$ (probability per unit time), so that $N$ cells give birth to $pN$ new cells per unit time, on the average. The cells can be located at different places in the tissue and may migrate upon multiplication. For such unrestricted growth, the differential equation is 

\begin{equation}
dN/dt = pN\:.
\label{exp1}
\end{equation}

Integrating Eq.~1, we obtain the number of cells at any time $t$ from the start of the growth process, given their initial value $N_0$, to be 

\begin{equation}
N(t)= N_0 \exp(p t)\;.
\label{exp2}
\end{equation}

This type of growth is unrestricted, the birth of a new cell has no effect on that of subsequent cells. This scenario probably holds true especially at the initial stages of growth for small values of malignant cell concentration. When the number $N$ of malignant cells grows within the tissue, higher concentrations will arise. We would like to argue that this process results in the spontaneous formation of macroscopic clusters of cells (which we will call tumors) which mop up almost all of the microscopic clusters, at a rather sharp transition point called the percolation transition (Stauffer and Aharony 1992), and which occurs here at a given concentration of sick $vs$ healthy cells within a given region.

At the percolation threshold (Stauffer and Aharony 1992), the clusters will be  fractal (Mandelbrot 1983).  For fractal clusters one will have the radius of gyration (a measure of the average radius) $R_g\sim N^{1/D}$, with $D < 3 $ being the fractal dimension, and $N$ is total number of sick cells within the tumor. It is also known that with more than one condensation site, one may have cluster-cluster aggregation which leads once again to fractal clusters (Kolb \etal 1983). Were the growth to continue at sites within this porous, fractal cluster, with the concentration of sick cells growing beyond the percolation threshold,  the resulting objects would be compact, with $D\to 3$.
%%with their radius of gyration  growing as $R_g\sim N^{1/3}$.  

It seems that simply due to spatial constraints (Kansal \etal 2000), or under-nourishment and under-oxygenation (hypoxia) of the interior (Kohandel \etal 2007), growth is confined essentially to the surface of the macroscopic clusters. For growth confined to the surface of the tumor, the kinetic equations should involve the number  $N_s$ of actively dividing cells in the surface layer, which scales as the surface area. Thus, $N_s\sim R_g^{D_s}$, with $D_s$ being the surface dimension. If the surface area can be said to grow as the radial derivative of the total mass, $D_s=D-1$. Then, $N_s \sim  R^{D_s} \sim N^{D_s/D}$.

The kinetic equation becomes, in the most general case where $D \le 3$, 
\begin{equation}
dN/dt \,=  \,p\, k_1 R_g^{D_s} = \,p\, k_2  N^{D_s/D}\,,
\label{ll1}
\end{equation}
where $k_1$ and $k_2$ are constants involving geometrical factors.  Integration yields, 
\begin{equation}
N \,= \,(c_1+ c_2 t)^u \,,
\label{ll2}
\end{equation}
where $u \,= \,D/(D-D_s)$, $c_1 = u\, N_0^{1/u}$, with $N_0$ having the same meaning as in Eq.~\ref{exp1}, and $c_2\,=\, p \,k_2 $.  Independently of  the exact value of the exponent, the behavior predicted by Eq.~\ref{ll2} certainly points to a time dependence that is a power law, for $c_1$ small enough. 

This analysis predicts a power-law growth, in all cases where the growth is essentially confined to the surface region of the tumors. Note that for ordinary compact objects with smooth surfaces, $D_s = 2$; for this value  Eq.~\ref{ll1} will give the expected growth rate for 
spheroidal tumors (Delsanto \etal 2004) and one gets $u\,=\,3$.  In the case that $D_s$ tends to $D$, meaning that all cells are free to multiply at the same rate, Eq.~\ref{ll1} will of course tend to Eq.~\ref{exp1}, with $k_2 \to 1$. For percolation clusters growing only along an outer shell (and not along the boundary or the ``hull" (Stauffer and Aharony 1992) of the highly convoluted cluster (Cao and Wong 1992)), one has precisely $u\,=\, D$, with $D\simeq 2.5$ in three dimensions (Stauffer and Aharony 1992).

A slightly different growth scenario is provided by the Eden model (Eden 1961, Jullien and Botet 1985, Wolf and Kert\'esz 1987, Family and Vicsek 1991, Meakin 1999). Growth starts from a seed or a surface, and one assumes that each cell within the surface layer has an equal probability to give rise to a daughter cell and the cells within the tumor multiply with a probability that is negligible, as recent results also show (Br\'u \etal 2003, Delsanto \etal 2004, Kansal \etal 2000). This growth rule is known to be in the same class of growth problems as the celebrated Kardar, Parisi, Zhang model (Kardar \etal 1986) of surface growth (Batchelor \etal 1998). The Eden model gives rise to compact tumors ($D\,=\,3$), with rough surfaces.  Although the surface is known to be self-affine, within a simplified fractal picture the surface area can be thought of as scaling with the average radius as $R_g^{D_s}$, where $D_s > 2$ is the fractal dimension of the surface.  The precise value of $D_s$ is under some debate. The width $w$ of the fluctuating surface region is known to scale as $w\sim R_g^\beta$ (Batchelor \etal 1998), with $\beta = 0.1$ (Kuennen and Wang 2008), yielding $D_s=2+\beta =2.1$. Alternatively $D_s$ may be estimated as $3-\alpha=2.3$, with the so called roughness exponent $\alpha = 0.7$ (Meakin 1999). Substituting these values in Eq.~\ref{ll2}, Eden growth predicts a growth exponent $u\,=\,3.3$, or $4.3$, depending upon the estimate for $D_s$.

Clearly, this discussion does not exclude the possibility  that upon further growth other mechanisms for inhibition or enhancement of the growth can come into play and that the growth regime will be further modified in qualitative agreement with the Gompertzian growth curve.

\subsection{Data analysis}

Fig.~\ref{log}(a) confirms an exponential growth of the initial PSA scores with a starting value of 0.0042 $\mu$g/l. The growth rate corresponds to $p=0.09\pm0.004$ per month. In everyday language, the PSA doubling time is about 8 months (the doubling time in months is here given by $(\ln 2)/p$); in other words, the PSA score increases more than three times within a year. Although the absolute PSA values  are much lower than the widely accepted threshold values for the recurrence of the prostate cancer, the growth rate is alarmingly fast. Yet, after about three years, the growth rate slows down and deviates from the exponential behavior. Nevertheless, an arbitrarily set threshold will be approached gradually.

Fig.~\ref{log}(b) is a log-log plot of the PSA values against the time. We observe that about two years after the operation there is a transition from unrestricted cell division at constant rate, characterized by exponential growth, to a power law signalling the formation of clusters and possibly the initiation of angiogenesis. The exponent of $u = 2.57\pm0.07$ is very close to the value predicted above, for percolation clusters with growth confined to an outer shell. We cannot, of course, rule out the Eden scenario based on just this result. It is possible that the effective number of surface cells is smaller than our naive estimation, thus giving rise to an overestimation of $D_s$, and therefore of $u$, in this case. Once the switch to power law growth occurs, signaling discrete, more or less compact tumors, we can estimate their size if the relation between the total number of cancerous cells and the PSA value is known. Assuming that the PSA level is linearly proportional to the total number of malignant cells, the PSA value and tumor size obtained from magnetic-resonance imaging (MRI) at the time of the operation provide our points of reference. In the present case, the prostate gland was imaged prior to its removal in August 2002 at a time when the PSA value was 8.5 $\mu$g/l. The gland had a diameter of less than  40~mm (Huber 2002). Taking into account the fact that the gland was about 50\% cancerous according to the post-operative biopsy, we deduced that a PSA level of 0.485~$\mu$g/l, measured in August 2008, would correspond to a tumor (consisting only of cancerous cells) of approximately 5~mm in diameter. Allowing for the possibility that not one, but two equal size compact clusters condense out of the scatter of individual cells, our naive calculation would give a diameter of approximately 4~mm for these tumors. 

It is highly significant that, digital rectal examinations by three independent experts (Brodmann and Riesterer 2008, Vollenweider 2008) as well as the MRI analysis (Hilfiker 2008)  performed in August 2008  revealed the presence of two masses of about 4~mm in agreement with our prediction. 

\section {Discussion}

Our result seems to be at variance with what is widely accepted as a ``universal law" (Schabel 1999, DeWys 1972, Norton \etal 1976, Torkington 1983, Iwata \etal 2000, Guiot \etal 2005, Delsanto \etal 2004, Br\'u \etal 2003, Kohandel \etal 2007) describing tumor growth, namely the ``Gompertzian" growth curve.  This curve is believed to  describe a gradual slowdown and eventual saturation in the growth of diverse populations from bacteria (Zwietering \etal 1990) to tumors and cell cultures (Norton \etal 1976) to natural populations in the wild.  It is commonly ascribed to a limiting ``carrier capacity"  of the environment, resulting in an inhibitory feedback of each new birth upon the rate of multiplication, due, say to depletion of nutrients (Zwietering \etal 1990)  or oxygen (Kohandel \etal 2007) or some other vitally important agent. Mathematically, this effect can be mimicked by introducing a rate of reproduction which decreases in time, as a function of the actual population.  This behavior was first mathematically described by Gompertz (1825). Although the precise mechanism of the slowdown in the case of tumor growth is still not totally clear, it is often characterized as obeying the Gompertzian growth law. This means an initial regime of exponential growth gradually slowing down and finally saturating to a constant value. (The final stage is to be observed in vitro, or in animal models.) One should remark that  plotting the data reported in Table 1 directly on the set marked with circles appearing in Fig.~3 of Norton \etal (1976) (after appropriately rescaling in the horizontal and vertical directions) results in an extremely good fit with the ``predicted" Gompertzian growth curve. This agreement suggests plotting similar sets of data on a log-log scale, to possibly reveal a sharp crossover from exponential to power law behavior in these sets as well.

\section {Conclusion}

It is well known (Kohandel \etal 2007, Norton \etal 1976, Guiot \etal 2005) that the initial stages of growth of many tumors, whether in vivo or in vitro, follow an exponential growth curve. This regime corresponds to dispersed particles that can grow without restriction. The crossover to a slower, power-law regime may be a manifestation of cluster growth, i.e., individual cells coagulate and become macroscopic structures with distinct surfaces. The clusters can be detected by some standard imaging techniques. Indeed, the presence of at least two small, but distinct masses was verified in the present case, after the onset of the power law growth regime.

Radiation therapy is not routinely applied after the surgery. The predictive power of our simple analysis, however, makes it highly worthwhile to closely monitor the PSA scores during the so-called wait-and-watch period no matter how low their absolute value is, as has been done in the present case, in order not to miss the optimum time window for post-operative therapy to increase the life expectancy of the cancer patient.

\section*{Acknowledgments}

ME thanks Dr. P. Vollenweider for his continued support and fruitful discussions during the active surveillance throughout the years before and after the prostate surgery. AE would like to thank M. Kardar for a helpful comment and  also acknowledge partial support by the Turkish Academy of Sciences.

\References
\item[] Batchelor M T, Henry B I and  Watt S D 1998 {\it Phys. Rev.} E {\bf 58} 4023
\item[] Bolla M \etal 2005 {\it Lancet} {\bf 366} 572
\item[] Brodmann S and Riesterer O 2008 {\it Medical Report} (Zurich: University Hospital)
\item[] Br\'u A, Albertos S, Subiza J L, Garcia-Asenjo J L and Br\'u I 2003 {\it Biophys. J.} {\bf 85} 2948
\item[] Cao Q and  Wong P 1992 {\it J. Phys.} A {\bf 25}, L69 It has been estimated by these authors that $99.8 \%$ of the sites on a percolation cluster are actually on the boundary, i.e., exposed to non-cluster sites.
\item[] Cozzarini C \etal 2004 {\it Int. J. Radiation Oncology Biol. Phys.} {\bf 59} 674 (2004)
\item[] Delsanto P P, Guiot C, Degiorgis P G,  Condat C A,  Mansury Y and Deisboek T S 2004 {\it App. Phys. Lett.} {\bf 85} 4225
\item[] DeWys W D 1972 {\it Cancer Research} {\bf 32} 374
\item[] Eden M 1961 {\it Proc. 4th Berkeley Symp. on Mathematical Statistics and Probability, Problems of Health} ed J Neyman  (Berkeley: University of California Press) p~223  cited by Batchelor M T, Henry B I and Watt S D 1998 {\it Phys. Rev.} E {\bf 58} 4023
\item[] Family F and Vicsek T ed {\it Dynamics of Fractal Surfaces} 1991 (Singapore: World Scientific, Singapore)
\item[] Gompertz B 1825 {\it Phil. Trans. R. Soc.} {\bf 115} 513
\item[] Guiot C, Degiorgis P G, Delsanto P P, Gabriele P and Deisboek T S 2005 {\it J. Theor. Biol.} {\bf 225} 147
\item[] Hilfiker P 2008 {\it Medical Report} (Zurich: MRI Bethanien)
\item[] Huber D 2002 {\it Medical Report} (Zurich: Klinik Hirslanden)
\item[] Iwata K, Kawasaki K and  Shigesada N 2000 {\it J. Theor. Biol.} {\bf 203} 177
\item[] Jullien R and Botet R 1985 {\it J. Phys.} A {\bf 18} 2279
\item[] Kansal A R, Torquato S,  Harsh IV G R, Chiocca E A and Deisboek T S 2000 {\it J. Theor. Biol.} {\bf 203} 367
\item[] Kardar M,  Parisi G and  Zhang Y-C 1986 {\it Phys. Rev. Lett.} {\bf 56}, 889
\item[] Kohandel M, Kardar M, Milosevic M and Sivaloganathan S 2007 {\it Phys. Med. Biol.} {\bf 52} 3665
\item[] Kolb M, Botet R and  Jullien R 1983 {\it Phys. Rev. Lett.} {\bf 51} 1123
\item[] Kuennen E W and  Wang C Y 2008 {\it J. Stat. Mech.} P05014
\item[] Mandelbrot B B 1983 {\it The Fractal Geometry of Nature} (New York: Macmillan)
\item[] Meakin P 1999 {\it Fractals, Scaling and Growth Far from Equilibrium} (New York: Wiley and Sons)
\item[] Norton L, Simon R,  Brereton H D and Bogden A E 1976 {\it Nature} {\bf 264} 542
\item[] Schabel Jr. F M 1969 {\it Cancer Research} {\bf 29} 2384
\item[] Stauffer D and  Aharony A 1992 {\it Introduction to Percolation Theory} (London: Taylor and Francis)
\item[] Torkington P 1983 {\it Bull. Math. Biol.} {\bf 45} 21
\item[] Vollenweider P 2008 {\it private communication}
\item[] Wolf D E and Kert\'esz J 1987 {\it Europhys. Lett.} {\bf 4} 651
\item[] Zwietering M H, Jongenburger I, Rombouts F M and van 't Riet K 1990 {\it Appl. Environmental Microbiol.} {\bf 56} 1875

\endrefs
\end{document}